\documentclass[reprint,amsmath,amssymb,aps,showkeys,prb]{revtex4-2}

\usepackage{float}
\usepackage{caption}
\usepackage{subcaption}
\usepackage{graphicx}
\usepackage{dcolumn}
\usepackage{bm}
\usepackage{rotating}

\begin{document}
	
\title{Calculation of Elastic Constants of UO$_2$ using the Hubbard-Corrected Density-Functional Theory DFT+U}
\author{Mahmoud Payami }
\email{mpayami@aeoi.org.ir}
\author{Samira Sheykhi}
\author{Mohammad-Reza Basaadat}
\affiliation{School of Physics \& Accelerators, Nuclear Science and Technology Research Institute, AEOI,\\ 
	P.~O.~Box~14395-836, Tehran, Iran}

\begin{abstract}
Uranium dioxide which is used as a fuel in light water nuclear reactors, is continually exposed
to radiation damage originated from the collision of high-energy particles. Accumulation of the
resulting defects gives rise to the evolution in the micro-structure of the fuel which in turn brings
about local tensions and strains in the fuel. One of the after effects due to evolution of
micro-structure is the swelling of fuel which can damage the fuel cladding and cause environmental contamination by leakage
of radioactive particles. Hence, it is vital to continually monitor the evolution of 
micro-structure and to analyze the changes in mechanical properties of the fuel. The study of
elastic constants and analysis of their behavior is very helpful in understanding the
mechanical properties of the fuel. In this research, using the Hubbard-corrected first-principles density-
functional theory method, we have calculated the elastic constants of the uranium
dioxide single crystal and compared the results with existing experimental data. In addition, using the Voigt, Reuss, and Hill models, we have estimated the mechanical properties for the poly-crystalline fresh fuel. The results show a very good agreement between the theory and experiment. Accordingly, we can reliably extend our method of calculations to the complicated system of irradiated fuel pellet, which
is in the form of a poly-crystal and hosts various defects.

\end{abstract}


\maketitle

\section{Introduction}\label{sec1}

\begin{figure}[ht]
	\centering
	\includegraphics[width=1.0\linewidth]{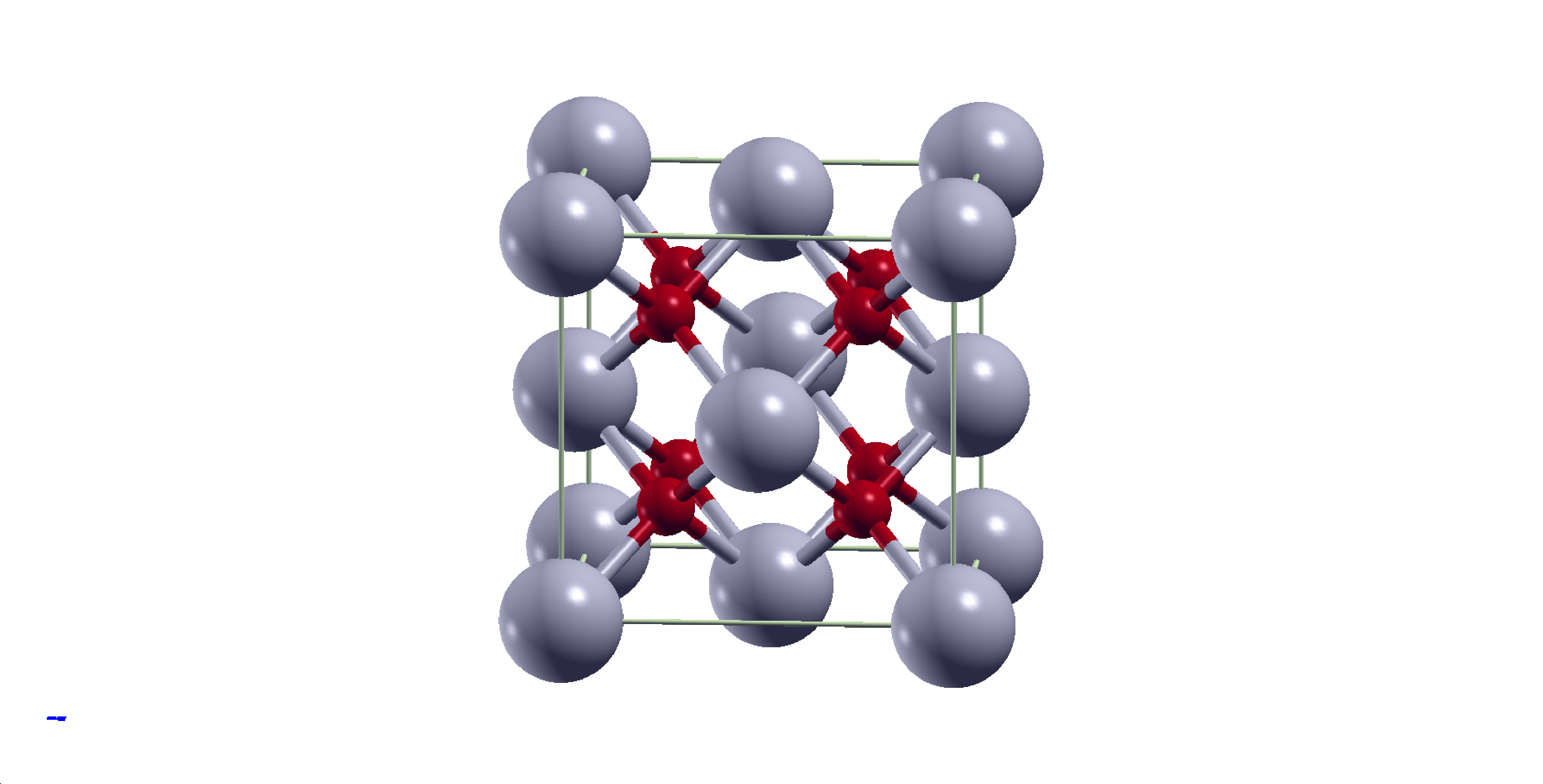}
	\caption{UO$_2$ crystal structure at low temperatures with cubic space group $Fm\bar{3}m$ (No. 225) with lattice constant of 5.47 \AA. Large grey and small red balls represent uranium and oxygen atoms, respectively.}\label{fig1}
\end{figure}

{\noindent \bf UO$_2$ Crystal}\\
UO$_2$ is used for years as a fuel with excellent performance in light water nuclear power reactors. The uranium dioxide crystal can be described with good accuracy using the cubic $Fm\bar{3}m$ space group (space group number 225) with a lattice constant 5.47\AA, as shown in Fig.~\ref{fig1}. Experimental investigations have revealed that UO$_2$ possesses an anti-ferromagnetic (AFM) crystal structure with a 3k-order at temperatures below $30~$K; para-magnetic form at higher $T$.

Extensive studies has shown that UO$_2$ is exceptionally resistant to radiation damage. However, due to persistent irradiation by high-energy particles or nuclear fission, crystal defects are created. The accumulation of defects lead to micro-structure changes in the fuel. Studies with swift heavy ion irradiation with energies about 70 MeV to 1 GeV, which mimic the radiation damage, were extensively performed \cite{wiss97,costantini07,thome09} in order to increase the understandings about radiation damage mechanisms. These experiments has shown that micro-structure changes lead to crystal lattice expansions and swelling of the fuel.\cite{tracy15,costantini-jap07}  
These processes lead to local stresses in the fuel which in turn give rise to local strains and thereof changes in the mechanical properties of the fuel. Therefore, the study of mechanical properties of the fuel finds itself as one of the most important topic in the field of nuclear materials. To investigate the mechanical properties of a material, usually the elastic constants and some derived quantities are analyzed. In this work, we have calculated the elastic constants and some derived quantities of UO$_2$ using the Hubbard-corrected density-functional theory, DFT+U\cite{payami-sheikhi23,payami24}. The results are in good agreement with experiment. 

\vspace{5pt}
{\noindent \bf DFT+U Method}\\
In the DFT \cite{hohenberg1964,kohn1965self} study of systems with highly correlated electrons, it has been shown that employing the local density (LDA) and generalized gradient (GGA) approximations for the exchange-correlation (XC), result in large self-interaction errors (SIE). The SIE lead to large delocalization of $d$ and $f$ orbitals which in turn result in the incorrect metallic behavior for Mott insulators. The relatively low-cost workaround for SIE is using the Hubbard model to correct the correlation energies of localized orbitals in the DFT energy functional. In this approach, which is called DFT+U, the energy functional is given by\cite{payami-sheikhi23,payami24}    

\begin{equation}\label{eq1}
	E_{\rm DFT+U}=E_{\rm DFT}[n({\bf r})] + E_{\rm Hub}[{n_m^{I\sigma}}]-E_{\rm dc}[n^{I\sigma}],
\end{equation}
in which $n({\bf r})$ is electron density, $n_m^{I\sigma}$ are occupation numbers of orbitals of atom at lattice site ${\bf R}_I$, and $n^{I\sigma}=\sum_m n_m^{I\sigma}$. The last term in right hand side of Eq.~(\ref{eq1}) is added to avoid double counting of interactions contained in the first and second terms.
The simplified rotationally invariant form of the correction is given by: 
\begin{equation}\label{eq2}
	E_{\rm U}[{n_{m m^\prime}^{I\sigma}}]\equiv E_{\rm Hub}-E_{\rm dc} =\sum_{I,\sigma}\frac{U^I}{2} {\rm Tr[{\bf n}^{\it I\sigma}({\bf 1} - {\bf n}^{\it I\sigma}})],
\end{equation}
in which ${\bf n}^{I\sigma}$ is the atomic occupation matrix. This correction, which is on-site correction, significantly improves the incorrect prediction of metallic behavior of Mott insulators and leads to correct prediction of insulator properties. The coefficients $U^I$ are called Hubbard on-site parameters. For a known material, these parameters may be empirically so tuned that the calculations give results in agreement with some experimental data and then, employing the tuned model, one is able to predict unknown other properties of that material.

In this work, using DFT+U, the elastic constants and some derived quantities has been calculated and shown to be in good agreement with experiment.

\section{Computational details}\label{sec2}
{\noindent \bf DFT+U Calculations}\\
The crystal structure of uranium dioxide is described with a cubic unit cell having 12 basis atoms as shown in Fig.~\ref{fig1}. To setup anti-ferromagnetic (AFM) structure for U atoms, we used the simple model in which the planes of U atoms alternate their spins when moving in $z$ direction, i.e., a 1-dimensional AFM. For the electron-ion interactions we have used scalar-relativistic ultra-soft pseudo-potentials (USPP) with PBEsol approximation for the XC. The valence configurations U($6s^2,\, 6p^6,\, 7s^2,\, 7p^0,\, 6d^1,\, 5f^3 $) and O($2s^2,\, 2p^4 $) were used in the USPP generation. All DFT+U calculations were based on the solution of the KS equations using the Quantum-ESPRESSO (QE) code package \cite{qe-2009,qe-2020}.  
 Kinetic energy cutoffs for the plane-wave expansions
were chosen as 90 and 720~Ry for the wave-functions and densities, respectively. The smearing method of Marzari-Vanderbilt for the occupations with a width of 0.01~Ry were used. 
For the Brillouin-zone integrations of geometry optimizations, a $6\times 6\times 6$ grid were used;  All geometries were fully optimized for total residual pressures on unit cells to within 0.5 kbar, and residual forces on atoms to within 10$^{-3}$~mRy/a.u. For Hubbard orbitals, we have used the atomic projection operators for the expansion of KS orbitals. The value of Hubbard parameter for the $5f$ orbital of uranium atoms is set to 2.45 eV \cite{payami23} which gives the lattice constant of optimized structure equal to the experimental value 5.47\AA.

{\noindent \bf Elastic Constants Calculations}\\
To calculate the elastic constants, we first fully optimize the geometry so that the forces on atoms and stress on crystal lattice vanish to a good accuracy. Then, we apply certain changes on the lattice vectors to make system strained and let the atoms relax to their new equilibrium positions. Now, for the strained system we calculate the stress tensor. Finally, having the stress and strain tensors at hand, we calculate the elastic constants. The second-rank stress and strain tensors, which are denoted respectively by $\sigma$ and $\varepsilon$, are related with a fourth-rank tensor as\cite{nye85}:

\begin{equation}\label{eq3}
	\sigma_{ij}=\sum_{kl}C_{ijkl}\varepsilon_{kl},
\end{equation}   
in which $i,j,k,l=x,y,z$ are Cartesian coordinates.
Both stress and strain tensors are symmetric and in the Voigt notation, under the mapping of indices $xx\mapsto 1 , yy\mapsto 2,zz\mapsto 3, yz\mapsto 4, xz\mapsto 5, xy\mapsto 6$, the above relation can be represented in matrix form as:

\begin{equation}\label{eq4}
	\begin{bmatrix}
		\sigma_1\\
		\sigma_2\\
		\sigma_3\\
		\sigma_4\\
		\sigma_5\\
		\sigma_6 
	\end{bmatrix}
	=
	\left[
	\begin{array}{cccccc}
		C_{11} & C_{12} & C_{13} & C_{14} & C_{15} & C_{16} \\
		C_{21} & C_{22} & C_{23} & C_{24} & C_{25} & C_{26} \\
		C_{31} & C_{32} & C_{33} & C_{34} & C_{35} & C_{36} \\
		C_{41} & C_{42} & C_{43} & C_{44} & C_{45} & C_{46} \\
		C_{51} & C_{52} & C_{53} & C_{54} & C_{55} & C_{56} \\
		C_{61} & C_{62} & C_{63} & C_{64} & C_{65} & C_{66} \\
	\end{array}
	\right]
	\begin{bmatrix}
		\varepsilon_1\\
		\varepsilon_2\\
		\varepsilon_3\\
		2\varepsilon_4\\
		2\varepsilon_5\\
		2\varepsilon_6\\
	\end{bmatrix}
\end{equation}  

The elastic tensor in this notation is a $6\times 6$ symmetric matrix with 21 independent components. If we denote the lattice vectors of optimized geometry by $\{\vec{a}_1, \vec{a}_2, \vec{a}_3\}$, then applying the following 6 independent transformation $F_i$ (one at a time) on optimized lattice vectors, we will have 6 strained states:
{\small 
\begin{equation}
	\begin{split}
	F_1=\left[\begin{array}{ccc}
		1+\delta & 0 & 0 \\
		0 & 1 & 0 \\
		0 & 0 & 1 \\
		\end{array}\right]
,
	F_2=\left[\begin{array}{ccc}
	1 & 0 & 0 \\
	0 & 1+\delta & 0 \\
	0 & 0 & 1 \\
    \end{array}\right]
,
	F_3=\left[\begin{array}{ccc}
	1 & 0 & 0 \\
	0 & 1 & 0 \\
	0 & 0 & 1+\delta \\
   \end{array}\right]
\\
	F_4=\left[\begin{array}{ccc}
	1 & 0 & 0 \\
	0 & 1 & \delta \\
	0 & 0 & 1 \\
   \end{array}\right]
,
	F_5=\left[\begin{array}{ccc}
	1 & 0 & \delta \\
	0 & 1 & 0 \\
	0 & 0 & 1 \\
   \end{array}\right]
,
	F_6=\left[\begin{array}{ccc}
	1 & \delta & 0 \\
	0 & 1 & 0 \\
	0 & 0 & 1 \\
   \end{array}\right]. \;\;\;\;\;\;\;\;\;\;\;\;\;\;\;\;\;\;
   \end{split}
\end{equation}
}
For each of the 6 strained states, we calculate the stress tensors for 4 different $\delta$ values, $\delta\in\{-0.01, -0.005, +0.005, +0.10 \}$, and then fit a straight line to extract the elastic constant from the linear coefficient.  

\section{Results and discussions}\label{sec3}
{\noindent \bf Geometry Optimization}
The calculations started using a 12 atom cubic cell (Fig.\ref{fig1}), the geometry fully optimized so that the forces on atoms and stress components on lattice vanish within a good accuracy. It should be stressed that in solving the KS equations for DFT+U method we always avoid the meta-stable states using the occupation-matrix-control (OMC) method\cite{payami-smcomc23}. The resulting equilibrium properties has been compared with experiment in Tab.\ref{tab1}.   

\begin{table}[ht]
	\caption{Equilibrium lattice constants, in \AA, total and absolute magnetizations, in Bohr-magneton per formula unit, equilibrium pressure on the lattice, in kilo-bar. }
	\begin{center}    
		\begin{tabular}{ ccccccc } 
			\hline \hline
			system  & results & $a=b$ & $c$ & tot-mag & abs-mag & P \\ \hline 
			UO$_2$  & present work & 5.465 & 5.478 &  0.00 & 2.14 & 0.01  \\ 
			UO$_2$  & experiment & 5.470 & 4.470 & - & - & -   \\ \hline 
		\end{tabular}\label{tab1}
	\end{center}
\end{table}

As is seen from Tab~\ref{tab1}, the system in equilibrium is slightly deformed from cubic symmetry in the $z$ direction. This originates from the fact that we have used a simplified 1-Dim AFM configuration in the $z$ direction.

\vspace{5pt} 
{\noindent \bf Elastic Constants}
To calculate the elastic constants, we applied different strains and calculated the corresponding stresses. Applying the strain operator $F_1$, we extract the 6 elastic constants: $C_{11}, C_{21}, C_{31}, C_{41}, C_{51}, C_{61} $.  
In addition, applying $F_2$, we obtain 6 other elastic constants: $C_{12}, C_{22}, C_{32}, C_{42}, C_{52}, C_{62} $, and so forth. One may benefit the symmetry relations to reduce the calculations, but we did not.
In Tab.~\ref{tab2}, we have listed the results and compared with the experimental values. As is seen, there is a very good agreement between theory and experiment. 

\begin{table*}
    \centering 
	\caption{Calculated elastic constants in GPa units. The experimental values\cite{wachtman65} are listed in parentheses. }
	\begin{tabular}{ ccccccc } 
		\hline \hline
		$C_{ij}$  & $j=1$ & $j=2$ & $j=3$ & $j=4$ & $j=5$ & $j=6$ \\ \hline 
		$i=1$  & $371.9 (395\pm 1)$ & $121.0 (121\pm 2)$ & $124.9$ & 0.0 & 0.0 & 0.0  \\ 
		$i=2$  & $121.1$ & $371.9$ & $124.9$ & 0.0 & 0.0 & 0.0  \\ 
		$i=3$  & $124.9$ & $124.8$ & $377.6$ & 0.0 & 0.0 & 0.0  \\ 
		$i=4$  & 0.0 & 0.0 & 0.0 & $72.5 (64.1\pm 1)$ & 0.0 & 0.0  \\ 
		$i=5$  & 0.0 & 0.0 & 0.0 & 0.0 & 72.5 & 0.0  \\ 
		$i=6$  & 0.0 & 0.0 & 0.0 & 0.0 & 0.0 & 70.6  \\ \hline 
	\end{tabular}\label{tab2}
\end{table*}

From symmetry considerations, the system has a tetragonal lattice in its optimized geometry with 1D
antiferromagnetic ordering. For such a system, the elastic constants matrix generally reflects tetragonal symmetry.
Non-zero $C_{ij}$ values align with this symmetry, and off-diagonal values like $C_{12}$
and $C_{13}$ show coupling between different directions.
Looking at diagonal terms $C_{11}$, $C_{22}$, and $C_{33}$, 
we see that $C_{11} = C_{22}$ = 371.9 GPa confirms in-plane isotropy in the $\vec{a}_1-\vec{a}_2$ plane due
to tetragonal symmetry.
The value $C_{33}$ = 377.6 GPa which is slightly higher, indicates greater stiffness along the
$\vec{a}_3$-axis compared to the $\vec{a}_1-\vec{a}_2$ plane.
Considering the off-diagonal terms $C_{12}$ abd $C_{13}$, 
$C_{12}$ = 121.0 GPa shows a strong coupling between the $\vec{a}_1$ and $\vec{a}_2$ axes.
Additionally, $C_{13}$ = 124.9 GPa and $C_{23}$ = 124.9 GPa suggest significant interaction
between the $\vec{a}_3$-axis and the other two $\vec{a}_1$ and $\vec{a}_2$ axes.
Now, focusing on shear moduli ($C_{44}, C_{55}, C_{66}$), 
$C_{44} = C_{55}$ = 72.5 GPa shows isotropy for shear deformation involving
the $\vec{a}_3$-axis.
$C_{66}$ = 70.6 GPa is slightly lower, indicating a small anisotropy for shear in
the $\vec{a}_1-\vec{a}_2$ plane.
For the coupling terms ($C_{21}, C_{31}, C_{32}$),
$C_{21}$ = 121.1 GPa, $C_{31} = C_{32}$ = 124.9 GPa suggest relatively uniform
coupling effects, which is consistent with tetragonal symmetry and
magnetic ordering effects.
The important physical insights from this study is:
i)-The high values of $C_{11}, C_{22}, C_{33}$ imply that UO$_2$ has significant stiffness,
making it robust against axial deformations;
ii)-The similar values for shear moduli ($C_{44}, C_{55}, C_{66}$) and their moderate
magnitudes point to balanced resistance to shear stresses, which is important
for mechanical stability of UO$_2$;
iii)-The symmetry in the elastic constants matrix aligns well with the tetragonal
lattice symmetry, confirming the correctness of our results.

\vspace{5pt}
{\noindent \bf Derived Elastic Properties}\\
A single crystal has a specific orientation, that is, its elastic response depends on direction (anisotropic).
On the other hand, a poly-crystalline material is made of many grains, each with random orientations.
So, for poly-crystals, the directional anisotropy averages out, that is, the overall bulk behavior matters.
Using the Voigt, Reuss, and Hill Averages\cite{hill1952} we estimate macroscopic (isotropic) elastic properties of poly-crystalline aggregates from anisotropic single-crystal elastic constants ($C_{ij}$).
In Voigt model, one assumes a uniform strain across grains, and it gives an upper bound for stiffness.
On the other hand, in Reuss model, one assumes a uniform stress across grains, and it gives a lower bound for stiffness. The arithmetic mean of Voigt and Reuss	gives the best average for isotropic poly-crystal.
To calculate the derived elastic properties, we need the compliance matrix $S$, which is the inverse of elastic constant matrix, $C$. Now, we list the definitions of each in Tab.~\ref{tab3}.

\begin{table*}[ht]
	\centering 
	\caption{Derived elastic properties. }
	\begin{tabular}{ cccc} 
		\hline \hline
		Property name  & Symbol & Formula & Value \\ \hline 
		Bulk Modulus (Voigt) & $B_V$ & $(1/9) * (C_{11} + C_{22} + C_{33}) + (2/9) * (C_{12} + C_{23} + C_{13})$ & 207.00 GPa  \\ 
		Bulk Modulus (Reuss) & $B_R$ & $1 / (S_{11} + S_{22} + S_{33} + 2*(S_{12} + S_{23} + S_{13}))$ & 206.97 GPa  \\
		Bulk Modulus (Hill Average) & $B_H$ & $ (B_V + B_R) / 2 $ & 206.99 GPa  \\
		Shear Modulus (Voigt) & $G_V$ & $(1/15) * (C_{11} + C_{22} + C_{33} - (C_{12} + C_{23} + C_{13})) + (1/5) * (C_{44} + C_{55} + C_{66})$  & 93.16 GPa  \\
		Shear Modulus (Reuss) & $G_R$ & $15 / (4*(S_{11} + S_{22} + S_{33}) - 4*(S_{12} + S_{23} + S_{13}) + 3*(S_{44} + S_{55} + S_{66}))$   & 86.60 GPa  \\
		Shear Modulus (Hill Average) & $G_H$  & $(G_V + G_R) / 2$  & 89.88 GPa   \\
		Young's Modulus (Voigt) & $E_V$  & $(9 * B_V * G_V) / (3 * B_V + G_V)$  & 243.02 GPa   \\
		Young's Modulus (Reuss) & $E_R$  & $(9 * B_R * G_R) / (3 * B_R + G_R)$  & 227.99 GPa   \\
		Isotropic Poisson's Ratio &  $\nu$  &  $(3 * B_H - 2 * G_H) / (6 * B_H + 2 * G_H)$ & 0.31    \\
		Universal Elastic Anisotropy Index & $A_U$ & $(5 * G_V / G_R) + (B_V / B_R) - 6$  & 0.38   \\  \hline 
	\end{tabular}\label{tab3}
\end{table*}

Bulk modulus represents resistance to uniform compression.
UO$_2$ is very resistant to volume change, with ~207 GPa being quite high.
Very close Voigt and Reuss values imply the material is nearly elastically isotropic in compression (i.e., behaves similarly in all directions).
Shear modulus tells us how well the material resists shape deformation (twisting or shearing).
Slight difference between Voigt and Reuss values imply some anisotropy in shear behavior.
Overall, the UO$_2$ material is relatively stiff, but not extremely rigid like some ceramics.
Young's modulus reflects the stiffness under uni-axial tension or compression.
Our calculated values are quite high, suggesting the material is strong and stiff, resisting stretching well.
The difference between Voigt and Reuss reflects some directional dependency, again hinting at slight anisotropy.
For the universal elastic anisotropy index, a value of 0 means perfect isotropy; the further from zero, the more anisotropic.
At 0.38, UO$_2$ material is almost isotropic, with minor directional variation in elastic response.
This is good for applications needing uniform mechanical behavior.
The Poisson's ratio, $\nu$ 
tells us how much the material contracts laterally when stretched.
Typical metals are around 0.30, brittle ceramics $<$ 0.25, and rubbery materials $>$ 0.4.
So, the value $\nu$=0.31 implies ductile-like behavior, not brittle.
Good balance of flexibility and strength.
To sum up results, UO$_2$ is stiff and incompressible material (high bulk and Young’s moduli).
It is nearly isotropic in elastic response (low anisotropy index).
It has moderate shear stiffness.
This material is suitable for load-bearing or pressure-resistant applications.

\section{Conclusions}\label{sec4}
Due to persistent irradiation of the UO$_2$ fuel by high-energy particles, the atoms are likely to displace from their lattice sites which results in crystal defect. In addition, the fission of heavy uranium atoms to lighter atoms makes the defects of the fuel more complicated. The accumulation of these defects gives rise to local stresses which in turn lead to local strains which is manifested as the swelling of fuel. The swelling may damage the fuel cladding and increase the risks of leakage of radioactive particles and environmental contamination. This point reveals the importance of the study of mechanical properties of the fuel. In this work, we have calculated the elastic constants of a fresh UO$_2$ single crystal and using the Voigt, Reuss, and Hill models, we have estimated the mechanical properties of fresh poly-crystalline UO$_2$ material. The method we have used for calculations is general and applicable when impurities are present. The good agreement between our results and experiment makes the method of calculation a reliable one for predicting the mechanical properties of the irradiated fuel.   

\section*{Acknowledgement}
This work is part of research program in School of Physics and Accelerators, NSTRI, AEOI.  

\section*{Data availability }
The raw or processed data required to reproduce these results can be shared with anybody interested upon 
sending an email to M. Payami.


\vspace*{2cm}
\section*{References}

\begin{thebibliography}{0}
	
\bibitem{wiss97} T. Wiss, H. Matzke, C. Trautmann, M. Toulemonde, S. Klaumünzer, {\it Nucl. Instrum. Methods} B {\bf 122} (3), 583–588 (1997).

\bibitem{costantini07} J. -M. Costantini, F. Beuneu, {\it Phys. Status Solidi} (c) {\bf 4} (3), 1258–1263 (2007).

\bibitem{thome09} L. Thomé, S. Moll, G. Sattonnay, et al., {\it J. Nucl. Mater.} {\bf 389} (2), 297–302 (2009).

\bibitem{tracy15} C. L. Tracy, M. Lang, J. M. Pray, et al., {\it Nat. Commun.} {\bf 6} (1), 6133 (2015).

\bibitem{costantini-jap07} J. -M. Costantini, C. Trautmann, L. Thomé, J. Jagielski, F. Beuneu, {\it J. Appl. Phys.} {\bf 101} (7), 073501(2007).

\bibitem{payami-sheikhi23} M. Payami, S. Sheykhi, and M.R. Basaadat, https://doi.org/10.48550/arXiv.2306.06266 (2023), and references therein.

\bibitem{payami24}  M. Payami, https://doi.org/10.48550/arXiv.2401.00864 (2024), and references therein.

\bibitem{hohenberg1964} P. Hohenberg and W. Kohn, {\it Phys. Rev.} {\bf 136}, B864 (1964).

\bibitem{kohn1965self} W. Kohn and L. J. Sham, {\it Phys. Rev.} {\bf 140}, A1133 (1965).

\bibitem{qe-2009} P. Giannozzi, S. Baroni, N. Bonini, et. al., {\it J. Phys.: Condensed Matt.} {\bf 21}, 395502 (2009).

\bibitem{qe-2020} P. Giannozzi, O. Baseggio, P. Bonfà, et. al., {\it J. Chem. Phys.} {\bf 152}, 154105 (2020).

\bibitem{payami23}  M. Payami, https://doi.org/10.48550/arXiv.2302.13381 (2023), and references therein.

\bibitem{nye85} J. F. Nye, {\it Physical Properties of Solids- Their Representation by Tensors and
Matrices}, OUP (1985).

\bibitem{payami-smcomc23} M. Payami, https://doi.org/10.47176/ijpr.23.3.11820, and references therein.

\bibitem{wachtman65} J. B. Wachtman Jr., M. L. Wheat, H. J. Anderson, J. L. Bates, {\it J. Nucl. Mat.}
	{\bf 16} (1), 39-41 (1965).

\bibitem{hill1952}  R Hill, {\it Proc. Phys. Soc. A} {\bf 65}, 349 (1952), https://doi.org/10.1088/0370-1298/65/5/307.

\end{thebibliography}
\end{document}